\documentstyle[aps,prl,preprint]{revtex}
\begin{document}
\title{On the contrasting spin dynamics of $La_{2-x}Sr_xCuO_4$,
$Nd_{2-x}Ce_xCuO_4$ and $YBa_2Cu_3O_{6+x}$ near half filling}
\author{Andrey V. Chubukov and Karen A. Musaelian}
\address{Department of Physics, University of Wisconsin - Madison,\\
1150 University Avenue, Madison, WI 53706\\
and\\
P.L. Kapitza Institute for Physical Problems\\
ul. Kosygina 2, Moscow, Russia 117334}
\date{June 23, 1994}
\maketitle
\begin{abstract}
We present simple calculations which show that
the incommensurability upon doping and the width of the
magnetically ordered phase in Mott-Hubbard insulators depend
strongly on the location of the hole/electron pockets in the Brillouin zone.
For $LaSrCuO$ systems,
we found the pockets at $(\pm \pi/2,\pm \pi/2)$, in which case the
corrections to the antiferromagnetic spin stiffness rapidly grow with doping
and destroy commensurate spin ordering already at a very small doping.
On the other hand, in $NdCeCuO$, the hole pockets are located at
$(\pi,0)$ and the symmetry related points, in which case the
 corrections to the stiffness scale linearly with the density of carriers and
do
not destroy commensurate spin ordering. For $YBCO$ systems, the situation is
less certain, but our results favor hole pockets at $(\pi/2,\pi/2)$.
We also briefly discuss the tendency towards phase separation.
\end{abstract}
\narrowtext

\section {introduction}

The intense interest in understanding the properties of high temperature
superconductors initiated theoretical research on the behavior of
antiferromagnetic insulators upon doping. The parent compounds of high-$T_c$
materials are well described as Heisenberg antiferromagnets. Upon hole doping,
long-range antiferromagnetism rapidly disappears and the systems eventually
become metallic superconductors.  The same transformation occurs in
electron-doped materials but at substantially larger doping concentrations.
The behavior of the antiferromagnetic insulator upon doping
attracted a lot of interest over the past few years  as the
exchange of antiferromagnetic paramagnons is at least one of the relevant
pairing interactions between holes~\cite{pines}.
 There are several fundamental issues
related to doped antiferromagnets, one of which is whether magnetic
correlations remain peaked at $Q_0 =(\pi,\pi)$ upon doping, or shift to
incommensurate momenta. Shraiman and Siggia first pointed out~\cite{ShSi}
 that if the  dispersion of vacancies has a
minimum at $(\pm \pi/2,\pm \pi/2)$, then
the dopants introduced into commensurate nearest-neighbor antiferromagnet
give rise to a
long-range dipolar distortion of the staggered magnetization which may lead to
a spiral spin configuration.

It has been recently argued~\cite{hyber,jap,gooding}
 that the values of hopping integrals in electron-and hole doped $214$
materials are
nearly the same; $t \sim 0.4ev,~t^{\prime} \sim -0.2t$, where $t^{\prime}$ is
the nearest-neighbor hopping. Despite this,
 the incommensurability upon doping has been found
only in $LaSrCuO$ compounds~\cite{aeppli}
while the dynamical structure factor in the doped $NdCeCuO$
remains peaked at $(\pi,\pi)$~\cite{gooding,el-dop}.
Moreover, experimentally,
long-range magnetic order in $LaSrCuO$,
disappears already at $3-4\%$ doping, while in
$NdCeCuO$ it survives up to $12\%$ doping.

The goal of the present paper is to show that the
 contrasting magnetic dynamics in the two 214 systems near half filling
is related to a different location of the hole pockets.
Namely, we will argue that
in $La-$based materials, the pockets are located at
$(\pm \pi/2,\pm \pi/2)$ while
in $Nd-$ based materials doped electrons occupy pockets centered around $(0,\pm
\pi)$ and $(\pm \pi,0)$. This, as we show below, gives rise to a completely
different spin dynamics in the two materials: the commensurate $(\pi,\pi)$
state rapidly becomes unstable in $LaSrCuO$, but survives in $NdCeCuO$.

 We will also discuss the location of the hole pockets in
 $YBa_2Cu_3O_{6+x}$ where the next-nearest neighbor hopping amplitude
is relatively large, $t^{\prime} \sim -0.5t$.
 Our results for the Hubbard model,
 show that for $t^{\prime}/t
\sim -0.5$ and $J/t \sim 0.4$, the mean-field hole dispersion is nearly
degenerate along $k_x =k_y$ and has a flat minimum at $(\pi,\pi)$.
However, self-energy corrections still favor pockets at $(\pi/2,\pi/2)$ and
are likely to overshadow the small difference between the
mean-field quasiparticle energies at $(\pi/2,\pi/2)$ and $(\pi,\pi)$.
In this situation,  the
 spin dynamics of $LaSrCuO$ and $YBaCuO$
near half-filling are nearly identical, and  differ only in the metallic
phase where the Fermi surface is large, and in case of $YBaCuO$, is centered
at $(\pi,\pi)$.
Notice, however,
that cluster calculations for the
$t-t^{\prime}-J$ model reported
 hole pockets at $(\pi,\pi)$ for
 the same ratios of parameters~\cite{gooding,dag}.
If it is actually the case for $YBaCuO$, the
spin dynamics right near
half-filling will be very similar to that in
the electron-doped materials (see below).

 The bulk of our consideration is presented in the
next section. We will first
briefly review the spin-density wave theory for the Hubbard model near
half-filling, then  will find the location of the hole pockets for the hole-
and electron-doped materials, and next show how the different location of the
hole pockets give rise to a contrasting magnetic behavior near half-filling.
 Finally, we
discuss the tendency towards domain wall
formation upon doping.
Our conclusions are presented in Sec.\ref{concl}.

\section{Hubbard model with next-nearest-neighbor hopping}

We consider the one-band Hubbard model given by
\begin{equation}
{\cal H} = -t\sum_{<i,j>} a^{\dagger}_{i,\sigma}a_{j\sigma} -
t^{\prime}\sum_{<i,j^{\prime}>} a^{\dagger}_{i,\sigma}a_{j^{\prime}\sigma} +
U \sum_i n_{\uparrow}n_{\downarrow}
\label{hub}
\end{equation}
Here $j$ and $j^{\prime}$ label the nearest and the next-nearest neighbors,
respectively, and $n = c^{\dagger}c$ is the particle density.
We will use the
spin-density-wave formalism~\cite{shrief}
 which, as has been shown in a number of
papers~\cite{shrief,bedell,john,S-T,chub_fren,chub_mus1,chub_mus2}, is a
good starting point for the calculations close to half-filling. Below we
mostly restrict with the simplest mean-field calculations.
This last restriction
can be formally justified if one extends the Hubbard model to a large
number of orbitals at a given site, $n_c =2S$,~\cite{haldane}
 and restrict with the
leading term in $1/S$ expansion. This mean-field theory is meaningless for
the nearest-neighbor Hubbard model because of the accidental degeneracy in
the hole spectrum which is lifted only by $1/S$ corrections (see the
discussion below).
However, the nonzero $t^{\prime}$ eliminates the accidental degeneracy
already at the mean-field level. In this situation, we expect
that the corrections to the mean-field
results renormalize the parameters of the model,
which will be important for our
analysis of $YBaCuO$, but do not give rise to any
new physics of the insulating phase.

\subsection{SDW theory at half-filling}

We now briefly discuss
the key points of the SDW formalism at half-filling. This formalism has been
applied  several times to the $t^{\prime} =0$ model. First, we
assume that at half-filling, the $2D$ Hubbard model has a commensurate
antiferromagnetic ground state. This implies that, e.g., $z$ component of
the spin-density operator, ${\vec S}(q) = (1/2) \sum_k
a^{\dagger}_{k+q, \alpha}
{\vec \sigma}_{\alpha,\beta} a_{k,\beta}$, has a nonzero expectation value at
$q =Q_0$. We then use the relation $\langle \sum_k a^{\dagger}_{k+Q,
\uparrow}a_{k,\uparrow} \rangle =
-\langle \sum_k a^{\prime}_{k+Q, \downarrow}a_{k,\downarrow}\rangle
 = \langle S_z \rangle$, to decouple the quartic term in (\ref{hub}).
 After decoupling, the
quadratic Hamiltonian takes the following form
\begin{eqnarray}
{\cal H_{MF}} &=& {\sum_k}^{\prime} \epsilon^{+}_k
(a^{\dagger}_{k\sigma}a_{k\sigma} +
a^{\dagger}_{k+Q_0\sigma}a_{k+Q_0\sigma}) + \nonumber \\
&& {\sum_k}^{\prime} \epsilon^{-}_k
(a^{\dagger}_{k\sigma}a_{k\sigma} -
a^{\dagger}_{k+Q_0\sigma}a_{k+Q_0\sigma}) - \nonumber \\
&& {\sum_k}^{\prime}  \Delta sgn(\sigma)~
(a^{\dagger}_{k\sigma}a_{k+Q_0\sigma} +
a^{\dagger}_{k+Q_0\sigma}a_{k\sigma})
\label{2}
\end{eqnarray}
Primes to the summation signs indicate that the summation is over the reduced
Brillouin zone. We introduced $ \Delta = U\langle S_z \rangle, ~
\epsilon_k = -2t(\cos k_x + \cos k_y) - 4t^{\prime} \cos k_x \cos k_y,~
\epsilon^{+} = (\epsilon_k + \epsilon_{k+Q_0})/2 =
 -4t^{\prime} \cos k_x \cos k_y,~
\epsilon^{-} = (\epsilon_k - \epsilon_{k+Q_0})/2
 = -2t(\cos k_x + \cos k_y)$. The next step is the
diagonalization of the quadratic form by a Bogolyubov transformation
\begin{eqnarray}
a_{k,\sigma} &=& u_k c_{k\sigma} + v_k d_{k\sigma} \nonumber \\
a_{k+Q_0,\sigma} &=& sgn(\sigma) (u_k d_{k\sigma} - v_k c_{k\sigma})
\label{uv}
\end{eqnarray}
Applying this transformation to (\ref{2}), we observe that the first term
with the density of quasiparticles, indeed, does not depend on $u_k$ and $v_k$,
because the transformation
conserves the total density. The Bogolyubov coefficients then
appear only in the last two terms which do not depend
on $t^{\prime}$. As a result, the expressions for $u_k$ and $v_k$ remain the
same as in the $t^{\prime} =0$ model~\cite{shrief}:
\begin{equation}
u_k = \left[ \frac{1}{2} \left(1 +
\frac{\epsilon^{-}_k}{E^{-}_k}\right)\right]^{\frac{1}{2}};~~
v_k = \left[ \frac{1}{2} \left(1 -
\frac{\epsilon^{-}_k}{E^{-}_k}\right)\right]^{\frac{1}{2}}
\end{equation}
where $E^{-}_k = \sqrt{\Delta^2 + (\epsilon^{-}_k)^2}$.

After the diagonalization, Eq.(\ref{2}) takes the form
\begin{equation}
{\cal H_{MF}} = {\sum_k}^{\prime} E^{c}_k c^{\dagger}_{k\sigma} c_{k\sigma} -
E^{d}_k d^{\dagger}_{k\sigma} d_{k\sigma}
\end{equation}
where
\begin{equation}
E^{c} = E^{-}_k + \epsilon^{+}_k, ~E^{d} = E^{-}_k -
\epsilon^{+}_k
\label{en}
\end{equation}
For $U \gg t$, which is implicit in our approach, we can expand under the
square root
and obtain $E^{c,d} = \Delta + J(\cos k_x + \cos k_y)^2 \mp 4 t^{\prime} \cos
k_x \cos k_y$, where $J = 4t^2/U$.
 We will refer to the quasiparticles described by $c$ and $d$
operators as conduction and valence fermions, respectively.
At half-filling, valence states are occupied and conduction states are empty.
Accordingly, the self-consistency condition on $\langle S_z \rangle$
 takes a simple form
\begin{equation}
\frac{1}{U} =  {\sum_k}^{\prime} \frac{1}{E^{-}_k}
\end{equation}
At large $U$, we obtain, as usual,  $\Delta = U/2$, or $\langle S_z\rangle
 \approx 1/2$.

\subsection{Finite density of holes}

We now discuss what happens at small but finite
doping when the chemical potential moves into the valence band.
First, we discuss the shape of the hole Fermi surface.
As we said above, at $t^{\prime}=0$,
the dispersion of valence fermions, Eq.(\ref{en}),
 is degenerate along the boundary of the
magnetic Brillouin zone ($k_x \pm k_y =\pm\pi$), where $E^{d} = \Delta$.
This degeneracy, however, is not related to any kind of symmetry and is removed
by self-energy corrections~\cite{S-T,chub_mus1}, with the result that
the actual band minima in the nearest-neighbor Hubbard model are
at $(\pm\pi/2,\pm\pi/2)$.
This agrees with the
numerical~\cite{Manous,gooding2,elbio} and
variational~\cite{Trug} studies of the Hubbard
 and $t-J$ models.

Our first observation for $t-t^{\prime}-U$ model is that at
finite $t^{\prime}$,
the degeneracy is removed at the mean-field level.
Indeed, a simple inspection of Eq.(\ref{en}) shows
that the mean-field dispersion has a minimum at $(\pm \pi/2,\pm \pi/2)$ if
$t^{\prime}$ is negative and smaller than $J$. For $|t^{\prime}|
>J$, the minimum of the hole dispersion is at $k =(\pi,\pi)$ (or $ (0,0)$).
Finally, if $t^{\prime}$ is
positive, which is probably not the case for cuprates,
the minimum of $E^{d}$ is at $(\pi,0)$ and symmetry related
points~\cite{gooding2}.
Now, for both 214 compounds,
$t^{\prime}$ is negative and relatively small:
 $|t^{\prime}| \sim 0.07 ev$, which  is
smaller
than the exchange integral $J\sim 0.13 ev$~\cite{raman,nmr,aepplisw}.
 Accordingly, we expect that upon doping, holes in $La-$ based
compounds form pockets around $(\pi/2,\pi/2)$ and the symmetry related points.
Near $(\pi/2,\pi/2)$, one can expand $E^{d}_k$ and obtain
\begin{equation}
E^{d}_k = \Delta + \frac{k^{2}_{\perp}}{2 m_{\perp}} +
\frac{k^{2}_{\parallel}}{2 m_{\parallel}}
\end{equation}
where  $m_{\perp} = 1/4(J - |t^{\prime}|),~m_{\parallel} =1/4|t^{\prime}|$.
Notice
that numerically, the effects
due to $t^{\prime}$, even for $|t^{\prime}/t| \sim 0.2$ are likely to dominate
over the effects due to self-energy corrections in nearest-neighbor model.
 Thus, for $t/J =2$,
the difference $\Delta E = E^{d} (\pi,0) - E^{d} (\pi/2,\pi/2)$ is $\Delta E =
4 |t^{\prime}| \sim 0.8t$ due to $t^{\prime}$
 and about $0.25 t$ due to quantum
fluctuations as was obtained in $1/S$ expansion for the Hubbard
model~\cite{chub_mus1} and in numerical~\cite{elbio} and
variational~\cite{Trug} calculations for the $t-J$ model.
The same is also true
for the inverse effective mass, $1/m_{\parallel}$:
for the same ratio $t/J$, the contribution to
$1/m_{\parallel}$ due to $t^{\prime}$ is about four times larger than due to
quantum fluctuations. Also notice that for the parameters chosen for $LaSrCuO$,
the two effective masses are roughly equal to each other, i.e., the Fermi
surface near half-filling is nearly  circular.

Consider now the electron-doped materials.
Under the electron doping, the chemical potential
moves into the conduction band. The energy of a conduction  fermion is
$E^{c} = \Delta + J(\cos k_x + \cos k_y)^{2} - 4t^{\prime} \cos k_x \cos
k_y$, i.e. it effectively has the sign of $t^{\prime}$ reversed compared to
the hole-doped materials. From the consideration above, we immediately conclude
that the minimum of the electron dispersion is at $(0,\pi)$ and the symmetry
related points. The same is, indeed, true also for
the selection due to quantum
fluctuations in the nearest-neighbor model.
 Expansion around the minima yields two equivalent effective
masses $m_{\perp} = m_{\parallel} = 1/4 |t^{\prime}|$.

Finally, consider the
electron dispersion in the $Y-$ based
hole-doped materials.
The spin dynamics of the overdoped $123$ systems was studied in a
number of papers by K. Levin, Q. Si and coauthors~\cite{si}.
They found that to fit the photoemission data for
$YBa_2Cu_3O_7$~\cite{pe},
one needs $t \sim 0.25-0.3ev$ and
relatively large next-nearest-neighbor hopping term
$t^{\prime} \sim -0.5t$~\cite{si}.
The values of the hopping integrals right near half-filling are not necessary
the same as in $YBa_2Cu_3O_7$
as the parameters of the
effective one-band Hubbard model derived from the underlying three-band model
generally depend on doping~\cite{comm}. We however simply assume that
 the values of $t$ and $t^{\prime}$ change little with
decreasing oxygen content. In this situation,
$|t^{\prime}|$ is very
close to $J$. This implies that the
mean-field hole dispersion, Eq(\ref{en}),
 is nearly degenerate along $k_x = k_y$: the band minima at $t =0.3 ev$
is at $(\pi,\pi)$,
but the quasiparticle energy at $(\pi/2,\pi/2)$ is only $0.08ev$ above.
To lift the near degeneracy, we calculated the leading
self-energy correction to the hole dispersion
in the expansion over the inverse
number of orbitals. The procedure is described in
some length in our earlier publication~\cite{chub_mus1}, and we do not discuss
it here.
We found, that fluctuations stabilize the minima at $(\pm \pi/2,\pm
\pi/2)$ up to much larger $t^{\prime}$ than in the mean-field theory.
Specifically,  the self-energy terms produce the energy difference,
$\Delta E \sim  0.88 \Delta$ (for $2S =1$), where
$\Delta E = E_{(\pi,\pi)} - E_{(\pi/2,\pi/2)}$.
For $t =0.3 ev$, we have
$\Delta =4t^2<S_z>/J \sim 0.9 ev$~\cite{optics} and hence
$\Delta E \sim 0.8 ev$.
 This implies that the actual critical value
of $|t^{\prime}|$ above which pockets are located at $(\pi,\pi)$,
 is about $0.32 ev$, which is substantially larger than
$|t^{\prime}| \sim 0.13-0.15 ev$ predicted for $YBaCuO$.

The hole pockets at $(\pm \pi/2,\pm \pi/2)$ near half-filling
are consistent with the results of photoemission
studies of the insulating $YBa_2Cu_3O_{6.3}$~\cite{arpes}. These studies
 have detected some spectral features
 which can be interpreted as the dispersion through Fermi surface, but {\it
only} close to the zone diagonal, i.e., near $(\pi/2,\pi/2)$.
Notice however that the closed Fermi surface near that point has not been
restored experimentally.

Another important point is that
the critical value of $t^{\prime}$ is indeed model dependent -
in the Hubbard-model calculations we found that it is larger than in $YBaCuO$,
but, as we already mentioned in the Introduction, small cluster
calculations for the  $t-t^{\prime}-J$ model found the minimum of hole
dispersion at $(\pi,\pi)$ for the same values of parameters as we used
{}~\cite{gooding}.

We further show how the different location of the pockets leads to a
contrasting
magnetic behavior near half-filling .

\subsection{Magnetic susceptibility}

In the SDW theory, the spin susceptibility is given by a ladder series
of bubble diagrams (Fig.\ref{bubble}).
One fermion in the bubble should be above the Fermi surface, and one
below. At half-filling, the only allowed
combination is one fermion from the conduction and one the from valence band.
 Away from half-filling, the Fermi level moves into the valence band, and
there are also bubbles with two valence fermions. The SDW expression for the
susceptibility has been derived earlier~\cite{shrief,chub_fren}, so we quote
only the result. In the static case, the total transverse susceptibility
$\chi^{+-} (q)$ is given by
\begin{equation}
\chi^{+-} (q) = \frac{\chi_0 (q)}{ 1 - U\chi_0 (q)}
\label{chitot}
\end{equation}
where
\begin{eqnarray}
\chi^{+-}_0 (q) &=& \frac{1}{2N} {\sum_{E^{d} >|\mu|}}^{\prime} \left[1 -
\frac{\epsilon^{-}_k \epsilon^{-}_{k+q} - \Delta^2}{E^{-}_k
E^{-}_{k+q}}\right] \left(\frac{1}{E^{c}_k + E^{d}_{k+q}} +
\frac{1}{E^{d}_k + E^{c}_{k+q}}\right) \nonumber \\
&&  \frac{1}{N}
{\sum_{\stackrel{E^{d}_{k+q} >|\mu|}{E^{d}_{k} <|\mu|}}}^{\prime}
 \left[1 +
\frac{\epsilon^{-}_k \epsilon^{-}_{k+q} - \Delta^2}{E^{-}_k
E^{-}_{k+q}}\right] \frac{1}{E^{d}_{k+q} - E^{d}_{k}}
\label{chi0}
\end{eqnarray}

Above we assumed that the system has a commensurate magnetic order. This
requires that the static spin susceptibility be non-negative
for all momenta (or, in other words, that all bosonic frequencies
 be real). Of special interest is the region near $q=Q_0 =(\pi,\pi)$
as $\chi^{-1}(q)$ turns to zero at $(\pi,\pi)$ in accordance with
the Goldstone theorem. Near this point, the
static susceptibility has the form~\cite{hydro}
\begin{equation}
\chi^{+-} (q) = \frac{2 N^{2}_0}{\rho_s (q - Q_0)^2}
\end{equation}
where $N_0$ is sublattice magnetization ($= 1/2$ in our mean-field approach),
and $\rho_s$ is the spin stiffness which should be positive.

Let us first consider half-filling.
Here only the first term contributes to $\chi_0$.
Performing an expansion in (\ref{chi0}) and substituting the result into
(\ref{chitot}), we obtain the ``classical'' spin wave result
\begin{equation}
\rho_s = \frac{1}{4}~J \left(1 - \frac{2 (t^{\prime})^2}{t^2}\right)
\label{rho}
\end{equation}
Clearly then, the commensurate $(\pi,\pi)$
state is stable at half-filling as long
as $\sqrt{2} |t^{\prime}| < t$. This condition,
though it may be  modified by quantum
fluctuations, is apparently satisfied in the $LaSrCuO$,
$NdCeCuO$ and $YBaCuO$ families.

We further consider the situation away from half-filling ($\delta \neq 0$).
 Now we also have a contribution from the second term which
involves only valence fermions.
Expanding in this term around $(\pi,\pi)$ and combining the result with
(\ref{rho}), we obtain
\begin{equation}
\rho_s (\delta) = \rho_s (0)~ (1 - z)
\label{renrho}
\end{equation}
where $\rho_s (\delta =0)$ is given by (\ref{rho}), and $z$ is
\begin{equation}
z = 4 U \frac{1}{N}\lim_{q \rightarrow 0}~
{\sum_{\stackrel{E^{d}_{k+q} >|\mu|}{E^{d}_{k} <|\mu|}}}^{\prime}
 \frac{\sin^{2} k_x}{E^{d}_{k+q} - E^{d}_{k}}
\label{z}
\end{equation}
At small concentration of holes, the condition $E^{d}_{k} <|\mu|$ implies that
the fermion with momentum $k$ is within the hole pocket.
For $La$ and $Y-$based materials,
these pockets are at $(\pm \pi/2,\pm \pi/2)$ where the
$\sin^{2} k$ factor in the numerator in (\ref{z}) is approximately one.
 Accordingly,
the summation over $k$ yields the uniform
Pauli susceptibility of free fermions, which
in two spatial dimensions
does not depend on the carrier concentration. Namely, for $z$ we obtain
\begin{equation}
z = 2U_{eff} \frac{\sqrt{m_{\perp} m_{\parallel}}}{\pi}
\end{equation}
For the case of $(\pi/2,\pi/2)$ pockets, $|t^{\prime}|\leq J$, so
that $\sqrt{m_{\perp} m_{\parallel}}$ scales as $1/J$. In the mean-field
theory, we also have $U_{eff} =U$ in which case
 $z \sim U/J$ is a large number, and
the spin stiffness immediately
changes sign upon doping which means that the commensurate $(\pi,\pi)$
antiferromagnetic state becomes unstable.
In more sophisticated calculations however, $U_{eff}$ appears
 different from $U$ because of the strong
self-energy and vertex corrections in the large-$U$ limit.
 In fact,
the self-consistent solution for $U_{eff}$ yields $U_{eff} \sim J$ at
$U \gg t$, and therefore
$z \sim O(1)$~\cite{ShSi,KLR,chub_fren}. Moreover, extremely close to
half-filling, $z$ vanishes logarithmically as  $z \sim 1/|\log{\delta}|$
because of the logarithmical singularity in the 2D scattering
amplitude~\cite{chub_fren}.
In any event, however, $z$ rapidly grows
with doping, and it is very likely that it quickly becomes larger than one in
which case the commensurate antiferromagnetic state is no longer stable. Note
that in the
mean-field approach we are using, this instability does not imply a disordering
transition, but rather a transformation into an incommensurate spin
configuration. The  equilibrium configuration at $z>1$ has
been discussed in our separate publication~\cite{chub_mus2}.

We now turn to the electron-doped systems. Here
 the hole pockets are formed around
$(0,\pi)$. Eq.(\ref{renrho}) and (\ref{z}) are still valid,
but the numerator in (\ref{z}) now
vanishes right at the
center of the pocket. Elementary calculations then show that because of the
$\sin^{2} k$ factor in (\ref{z}), $z$ scales {\it
linearly} with doping concentration, and hence at small doping,
$\rho_s$ acquires only a small correction $O(\delta)$. Clearly then,
antiferromagnetism at $(\pi,\pi)$ survives in the presence of a small density
of
electrons.  This explains why
$Nd_{2-x}Ce_xCuO_4$ remains commensurate all the way down to the
paramagnetic phase.

We now discuss the width of the magnetically ordered phase.
Within the present mean-field (or large $S$)
approach, the on-site magnetization is nearly
equal to its nominal value, and
 the rapid decrease in the stiffness
in the hole doped 214 materials is
not accompanied by the rapid decrease in the order parameter. In other words,
the mean-field theory predicts that the system first becomes incommensurate
and only then loses long-range order. There are, however,
numerous experimental reasons to believe that the disordering transitions
in weakly doped high-$T_c$ materials are in the universality class of the
nonlinear sigma model with the dynamical exponent $\bar{z}=1$~~\cite{csy}
 (the most direct
evidence is the observed linear behavior of the uniform susceptibility).
This implies than in a more adequate model,
 the decrease in $\rho_s$ must eventually
lead to the decrease in the sublattice magnetization such that both
quantities vanish simultaneously. This is what has been found by
Sachdev~\cite{Sub}
in the self-consistent large-N study of the Shraiman-Siggia model~\cite{ShSi}
in some range of the coupling constant values. In another
 range, he found
an incommensurate transition within the ordered phase like in our approach.
It is essential however, that the two scenarios
 differ primarily in the behavior of the
sublattice magnetization with doping, while the doping dependence of the spin
stiffness is nearly the same in both cases. In particular, for all values of
the coupling constant in the Shraiman-Siggia model (where pockets are at
$(\pi/2,\pi/2)$), the stiffness undergoes a
rapid, nearly step-like, downturn renormalization under hole doping.
We can, therefore, expect that the larger are the
corrections to the stiffness at low
doping (even if they are obtained in large-$S$ expansion, as in our approach)
 the smaller is the actual region of the magnetically ordered phase.
 Thus the width of the magnetically ordered phase in $NdCeCuO$ should
be much larger than in $LaSrCuO$. This is
consistent with the experimental observation that
magnetic order in
$NdCeCuO$  survives up to much larger doping concentrations than in
 $LaSrCuO$. A similar, though somewhat different,
explanation of the difference of the magnetic phase diagrams of the two
214 compounds, based on the idea of localized
electrons in $NdCeCuO$ and mobile holes in $LaSrCuO$, was
presented in~\cite{gooding}.

Finally, we notice that if the actual ratio of $t^{\prime}/t$ is such that
 the pockets in $YBaCuO$ at low doping
are located at $(\pi,\pi)$, then the corrections to the stiffness scale
linearly with $x$ by exactly the same reasons as in the
electron-doped compounds,
and the commensurate $(\pi,\pi)$ configuration survives the hole doping.

\subsection{Phase separation}

 In SDW approach, we can also consider stability of the $(\pi,\pi)$ phase at
small doping against the formation of domain walls~\cite{shulz,separ}.
This stability requires the longitudinal spin susceptibility to be positive.
 For the commensurate spin ordering, longitudinal spin fluctuations are
always decoupled  from transverse spin fluctuations, but at finite doping,
 they are coupled to charge fluctuations.
The total static uniform susceptibility $\chi^{zz}$ can be
obtained by straightforward manipulations starting
from Eq.(52)-(54) in~\cite{chub_fren}:
\begin{equation}
\chi^{zz} \approx \frac{2 \chi^{\rm Pauli}}{1 - 8J \chi^{\rm Pauli}}
\end{equation}
where $\chi^{\rm Pauli} = \sqrt{m_{\perp} m_{\parallel}}/2\pi$ is the
Pauli-like
susceptibility of doped carriers.
If this susceptibility is larger than $1/8J$,
the total longitudinal susceptibility becomes negative which signals the
formation of domain walls.

We calculated effective masses  with
 self-energy corrections for all three types of materials,
and found that for $|t^{\prime}| \sim J/2$,
the tendency towards phase separation is nearly the same in the
two 214 compounds (the Pauli susceptibility is slightly larger in $LaSrCuO$),
 and is much weaker in $YBaCuO$
where the Pauli susceptibility is about two
times smaller.
At the same time, we found that
 the denominator in $\chi^{zz}$ is positive in 214 materials,
i.e., there is a stability against domain wall formation
 immediately away from half-filling.
These results are consistent with the
 numerical analysis in Ref.\cite{gooding}. The latter paper also
 points to a possibility
of a two-dimensional
 phase separation in the electron-doped materials which we didn't
study.

\section{conclusions}
\label{concl}

To summarize, in this paper we presented simple calculations which show that
the stability of the commensurate antiferromagnetic state in
 the Mott-Hubbard insulators depends
strongly on the location of the hole/electron pockets in the Brillouin zone.
For $LaSrCuO$, we found pockets at $(\pm \pi/2,\pm \pi/2)$. The corrections to
the antiferromagnetic spin stiffness from the occupied hole states within
these pockets rapidly grow with the
carrier concentration and are likely to make stiffness negative,
i.e., destroy commensurate spin ordering, already at a very small doping.
On the other hand, in $NdCeCuO$, we found
that mobile electrons form pockets at $(0,\pi)$ and the symmetry related
points,
 in which case the
 corrections to the stiffness scale linearly with the density of carriers and
do
not destroy the commensurate spin ordering. We argued that the different
behavior
of stiffnesses is responsible for the experimentally observed difference in the
widths of the magnetically ordered phases in the two 214 compounds.
 These results compliment  the arguments
and numerical analysis in Ref\cite{gooding}.

We also discussed the hole dispersion in $YBaCuO$ and found that for the
value of $t^{\prime}$ used to fit the photoemission data,
band minima are
likely to remain at $(\pm\pi/2,\pm\pi/2)$ though the quasiparticle
energy at $(\pi,\pi)$ is only slightly larger.
 This implies that the magnetic
properties of $LaSrCuO$ and $YBaCuO$ are identical right near half-filling.
 At the first glance, this result
seems strange as the hole pockets at $(\pi/2,\pi/2)$ apparently
lead to incommensurability
which has been observed in neutron scattering experiments
only in $LaSrCuO$~\cite{aeppli,tranq}.
However, these experiments were made only
deep in the metallic phase when the Fermi surface is large and, in $YBaCuO$, is
centered around $(\pi,\pi)$. Recently, we
 considered~\cite{chub_mus2} the situation when
spin stiffness becomes negative while holes still occupy pockets at
$(\pi/2,\pi/2)$. We found that
the equilibrium static spin configuration is not the planar spiral
Shraiman-Siggia phase in which susceptibility is peaked at incommensurate
momentum,
 but rather a noncoplanar configuration which very much resembles
the $(\pi,\pi)$ state and differs from it
only in the existence of {\it small}
transverse spiral component of the order parameter, $S_{\perp}
\sim O(\sqrt{x})$. For this configuration, the
susceptibility still has a dominant peak at
 $(\pi,\pi)$. In other words, the spin structure adjustes to
the negative stiffness of the $(\pi,\pi)$ state in such a way that
the peak position of the susceptibility does not change as long as
hole occupy pockets at $(\pi/2,\pi/2)$.
Notice that this is consistent with the RPA-like analysis by Si
et al in the metallic phase as
very close to the magnetic transition they
 found the maximum in the susceptibility
at $(\pi,\pi)$ for both types of hole-doped materials.
The transformation of the
hole Fermi surface with increasing doping content from
small to large one,
and the related change in magnetic susceptibility still need to be studied.

\section{Acknowledgements}

It is our pleasure to thank E. Dagotto, D. Frenkel, R. Gooding, R. Joynt,
  and Q. Si
for useful conversations. We are also grateful to R. Gooding and E. Dagotto
for sending us
copies of Ref\cite{gooding} and~\cite{elbio} prior to publication.
The research was supported in part by the Graduate School at the University
of Wisconsin-Madison and Electric Power Research Institute.

\begin{figure}
\caption{The RPA series for the total static transverse susceptibility.
 The first term represents the
simple bubble, which is the building block of the ladder.
Solid and dashed lines denote valence and conduction fermions, respectively. At
half filling, only bubbles which contain one valence and one
conduction fermion
contribute to transverse susceptibility.}
\label{bubble}
\end{figure}


\begin{references}

\bibitem{pines} see e.g., P. Monthoux and D. Pines, Phys. Rev. B {\bf 47}, 6069
(1993) and references therein

\bibitem{ShSi} B. I. Shraiman and E. D. Siggia, Phys. Rev. Lett.
{\bf 60}, 740 (1988); {\bf 61}, 467 (1988).

\bibitem{hyber} M.S. Hybertsen  et al, Phys. Rev. B {\bf 41}, 11068 (1990).

\bibitem{jap} T. Tohyama and S. Maekawa, Phys. Rev B {\bf 49}, 3596 (1994).

\bibitem{gooding} R.G. Gooding, K.J.E. Vos and P.W. Leung, unpublished.

\bibitem{aeppli} S-W Cheong et al, Phys. Rev. Lett {\bf 67}, 1791 (1991).

\bibitem{el-dop} Y. Tokura, H. Takagi and S. Uchida, Nature (London), {\bf
337}, 345 (1989).

\bibitem{pe} R. Liu et al, Phys. Rev. B {\bf 45}, 5614 (1992).

\bibitem{si} Q. Si, Y. Zha, K. Levin and J.P. Lu, Phys. Rev. B {\bf 47}, 9055
(1993) and references therein.

\bibitem{tranq} J.M. Tranquada et al, Phys. Rev. B {\bf 46}, 5561 (1992).

\bibitem{dag}  E. Gagliano, S. Bacci and E. Dagotto
 Phys. Rev B {\bf 42}, 6222 (1990)

\bibitem{shrief} J.R.Schrieffer, X.G.Wen, and S.C.Zhang,
\prb {\bf 39}, 11663 (1989).

\bibitem{bedell} H.Monien and K.S.Bedell, \prb {\bf 45}, 3164 (1992).

\bibitem{john} S. John and P. Voruganti, \prb {\bf 43}, 10815 (1991);
S. John and P. Voruganti, and W.Goff, \prb {\bf 43}, 13365 (1991).

\bibitem{S-T} A. Singh and Z. Te\v{s}anovi\'{c}, \prb {\bf 41}, 11457 (1990);
G. Vignale and M.R. Hedayati, Phys. Rev. B {\bf 42}, 786 (1990).

\bibitem{chub_fren} A.V.Chubukov and D.M.Frenkel, \prb {\bf 46}, 11884 (1992).

\bibitem{chub_mus1} A.V. Chubukov and K. Musaelian, \prb, {\bf 50} ,... (1994).

\bibitem{chub_mus2} A.V. Chubukov and K. Musaelian, unpublished

\bibitem{haldane} I.Affleck and F.D.M.Haldane, \prb {\bf 36}, 5291 (1987).

\bibitem{gooding2} R.G. Gooding, K.J.E. Vos and P.W. Leung,
Phys. Rev. B {\bf 49}, 4119 (1994)

\bibitem{raman} see e.g., R.R.P. Singh, Comments Cond. Mat. Phys., Vol.15, No.
4, pp.241-257 (1991)

\bibitem{nmr} T. Imai et al, Phys. Rev. Lett {\bf 70}, 10002 (1993)

\bibitem{aepplisw} S.M. Hayden et al, Phys. Rev. Lett {\bf 66}, 821 (1991).

\bibitem{Manous} E. Dagotto, R. Joynt, A. Moreo, S. Bacci and E. Gagliano,
Phys.
Rev B {\bf 41}, 9049 (1990); M. Boninsegni and E. Manousakis,
Phys. Rev. B {\bf 43}, 10353 (1991).

\bibitem{elbio} E. Dagotto, A. Nazarenko and M. Boninsegni,
unpublished.

\bibitem{Trug} S. Trugman, Phys. Rev B {\bf 37}, 1597 (1988);
S. Sachdev, Phys. Rev B {\bf 39}, 12232 (1989).

\bibitem{comm}  The
relation between  $x$ and $\delta$ in $YBaCuO$
is also not firmly established and
currently is subject to
interpretation (see e.g.,  A. Sokol and D.
Pines, Phys. Rev. Lett {\bf 71}, 2813 (1993)).

\bibitem{optics} Notice that $2\Delta \sim 1.8 ev$ is consistent with
the position of the peak in the optical absorption, see e.g., D.B. Tanner and
T. Timusk, in {\it {Physical Properties of High Temperature Superconductors}},
edt. by D.M. Ginzberg (World Scientific, Singapore, 1992).

\bibitem{arpes} R. Liu et al, Phys. Rev. B {\bf 46}, 11056 (1992).

\bibitem{hydro} B.I.Halperin and P.C.Hohenberg, Phys. Rev.
{\bf 188}, 898 (1969);
D.Forster, {\em Hydrodynamic Fluctuations, Broken Symmetry, and
Correlation Functions}
(Benjamin/Cummings, Reading, MA, 1975)

\bibitem{KLR} C. L. Kane, P. A. Lee, and N. Read, \prb {\bf 39}, 6880 (1989).

\bibitem{csy} A. Chubukov, S. Sachdev and J. Ye, Phys. Rev. B {\bf 49}, xxx
(1994).

\bibitem{Sub} S. Sachdev, Phys. Rev. B {\bf 49}, xxx (1994).

\bibitem{shulz} H.J. Shulz, \prl, {\bf 64}, 1445 (1990).

\bibitem{separ} V.J. Emery and S.A. Kivelson, Physica C {\bf 209}, 597 (1993).

\end{references}
\end{document}